\documentclass[sigconf]{acmart}

\usepackage{graphicx}
\usepackage{hyperref}

\usepackage{balance}  
\usepackage{graphics} 
\usepackage{times}    
\usepackage{url}      
\usepackage{graphicx}
\usepackage{tabularx}
\usepackage{float}
\usepackage{color}
\usepackage{url}
\usepackage[noend]{algpseudocode}
\usepackage{algorithm}
\usepackage{verbatim}
\usepackage{mathtools}
\usepackage{caption}
\usepackage{subcaption}

\usepackage{amsthm}
\usepackage{thmtools}
\usepackage{xspace}
\usepackage{multirow}
\usepackage[capitalise]{cleveref}
\usepackage{amssymb}
\usepackage{amsmath}

\hypersetup{colorlinks=true,allcolors=blue}

\newcommand{\ea}{{\em et al.}\xspace}

\algdef{SE}[SUBALG]{Indent}{EndIndent}{}{\algorithmicend\ }%
\algtext*{Indent}
\algtext*{EndIndent}

\usepackage{hyperref}

\newcommand{\nameNNST}{{$NN_{ST}$}\xspace}
\newcommand{\nameNNInit}{{$NN_{init}$}\xspace}
\newcommand{\nameNNObserved}{{$NN_{Observed}$}\xspace}

\setcopyright{none}

\begin{document}

\title{Learning State Representations for Query Optimization with Deep Reinforcement Learning}

\author{
Jennifer Ortiz$^\dagger$, Magdalena Balazinska$^\dagger$, Johannes Gehrke$^\ddagger$, S. Sathiya Keerthi$^+$}
\affiliation{University of Washington$^\dagger$, Microsoft$^\ddagger$, Criteo Research$^+$}

\begin{sloppypar}
\begin{abstract}
Deep reinforcement learning is quickly changing the field of artificial intelligence. These models are able to capture a high level understanding of their environment, enabling them to learn difficult dynamic tasks in a variety of domains. In the database field, query optimization remains a difficult problem. Our goal in this work is to explore the capabilities of deep reinforcement learning in the context of query optimization. At each state, we build queries incrementally and encode properties of subqueries through a learned representation. The challenge here lies in the formation of the state transition function, which defines how the current subquery state combines with the next query operation (action) to yield the next state. As a first step in this direction, we focus the state representation problem and the formation of the state transition function.  We describe our approach and show preliminary results. We further discuss how we can use the state representation to improve query optimization using reinforcement learning. 
\end{abstract}


\maketitle

 \section{Introduction}\label{sec:deeplearning}
Query optimization is not a solved problem, and existing database management systems (DBMSs) still choose poor execution plans for some queries~\cite{Leis:15}. Because query optimization must be efficient in time and resources, existing DBMSs implement a key step of cardinality estimation by making simplifying assumptions about the data (e.g., inclusion principle, uniformity or independence assumptions)~\cite{Kiefer:17, Leis:15}. Additionally, while
research papers have shown their benefits, optimizers shy away from using multidimensional histograms and sampling due to the increased overhead and complexity they bring~\cite{Eavis:07,Wu:16}. As a result, in data sets with correlations and non-uniform data distributions, cardinality estimation errors are frequent, leading to sub-optimal plan selections~\cite{Leis:17}. 

\begin{figure}[t]
    \centering
    \includegraphics[width=.9\linewidth]{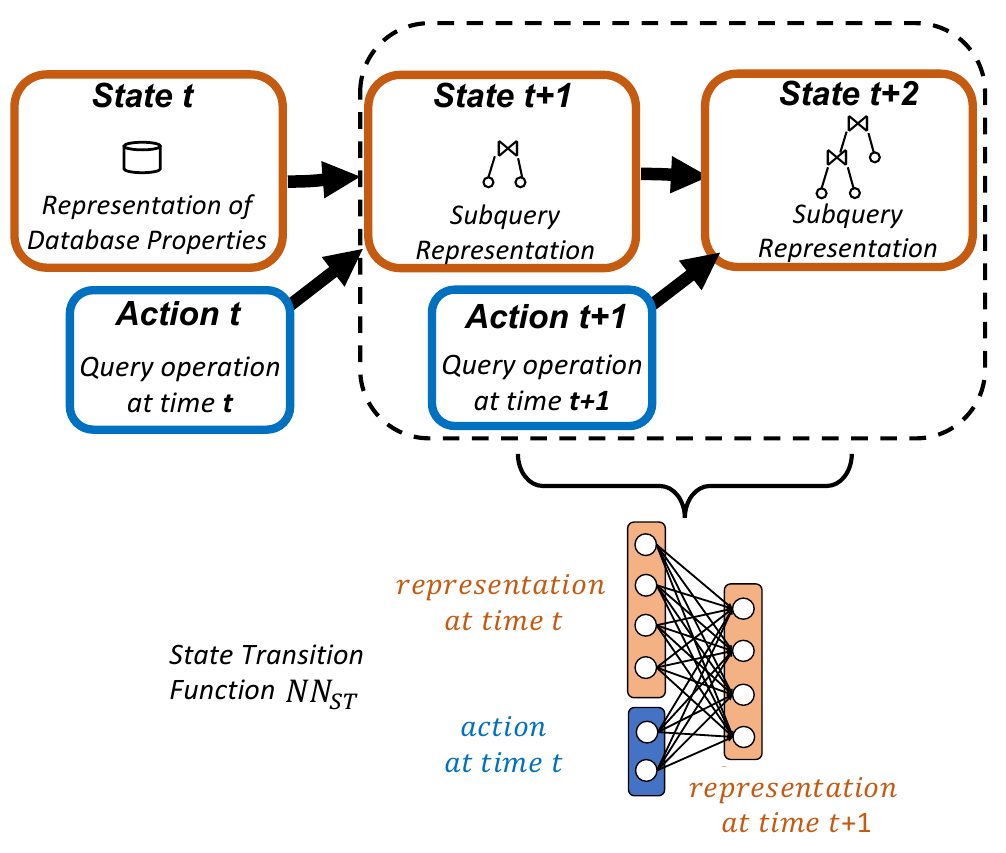}
    \caption{\textbf{State Representation}: Given a database and a query, we use a neural network to generate subquery representation for each state. This representation can serve for cardinality estimation and more importantly, for building optimal query plans using reinforcement learning.}
    \label{fig:overview}
\end{figure}

Recently, thanks to dropping hardware costs and growing datasets available for training, \textit{deep  learning} has successfully been applied to solving computationally intensive learning tasks in other domains. The advantage of these type of models comes from their ability to learn unique patterns and features of the data that are difficult to manually find or design~\cite{Goodfellow:16}. 

In this paper, we explore the idea of training a deep learning model to predict query cardinalities. Instead of relying entirely on basic statistics to estimate costs, we train a model to automatically learn important properties of the data to more accurately infer these estimates. This representation can result in better predictions than using hand-designed feature selection~\cite{Goodfellow:16}. They can automatically learn to retain distinguishing properties of the data~\cite{Lesort:18}, which in turn can help estimate cardinalities. As of today, there are few studies that have used deep learning techniques to solve database problems, although some have started to raise awareness for the potential of this method in our field~\cite{Wang:16_2}. Now is the time to explore this space, since we have the computational capabilities to run these models. A key challenge of this approach is how to represent the queries and data in the model. If we build a model that takes as input a feature vector representing a complex query on a database, this model would essentially need to learn to predict cardinalities for \textit{all} possible queries that could run on the data. Such a model would be complex and would require impractical numbers of training examples. 

To address the above problem, as a first contribution (\autoref{sec:state_representation}), we develop an approach that trains a deep learning model that learns to \textit{incrementally} generate a succinct representation of each subquery's intermediate results: The model takes as input a subquery and a new operation to predict the resulting subquery's properties. These properties can serve to derive the subquery's cardinality.

As a second  contribution (\autoref{sec:reinforcement_learning}), we present our initial approach to using these representation to improve query plan enumeration through reinforcement learning. Reinforcement learning is a general purpose framework used for decision-making in contexts where a system must make step-by-step decisions to reach an ultimate goal and collect a reward. In our case, we propose to use this approach to incrementally build a query plan by modeling it as a Markov process, where each decision is based on the properties of each state.

In ~\autoref{fig:overview}, we show an example that forms the basis of our deep reinforcement learning approach. Given a query and a database, the model incrementally executes a query plan through a series of state transitions. In the initial state $t$ in the figure, the system begins with a representation of the entire database. Given an action selected using reinforcement learning, the model transitions to a new state at $t+1$. Each action represents a query operation and each state captures a representation of the subquery's intermediate results. We train a state transition function (a neural network), \nameNNST, to generate this representation. \nameNNST is a recursive function that takes as input a previous subquery representation as well as an action at time $t$, to produce the subquery representation for time $t+1$.

Let us now motivate the setup that is laid out in~\autoref{fig:overview}. 
Consider the dynamics of a query plan that is executed one operation (action) at a time. At any stage $t$ of the query plan execution, let's say a subquery has been executed; let $h_t$, the state at $t$ be represented by an $n$-dimensional real vector. Applying the next action, $a_t$ to this current database state leads to the next state, $h_{t+1}$. The mapping, $NN_{ST}:(h_t,a_t)\rightarrow h_{t+1}$ is called the state transition function.

In most applications of reinforcement learning, the state as well as the state transition function are known. For example, in the game of Go, each possible board position is a state and the process of moving from one board position to the next (the transition) is well-defined. 

\begin{figure}[t]
    \centering
    \includegraphics[width=0.5\linewidth]{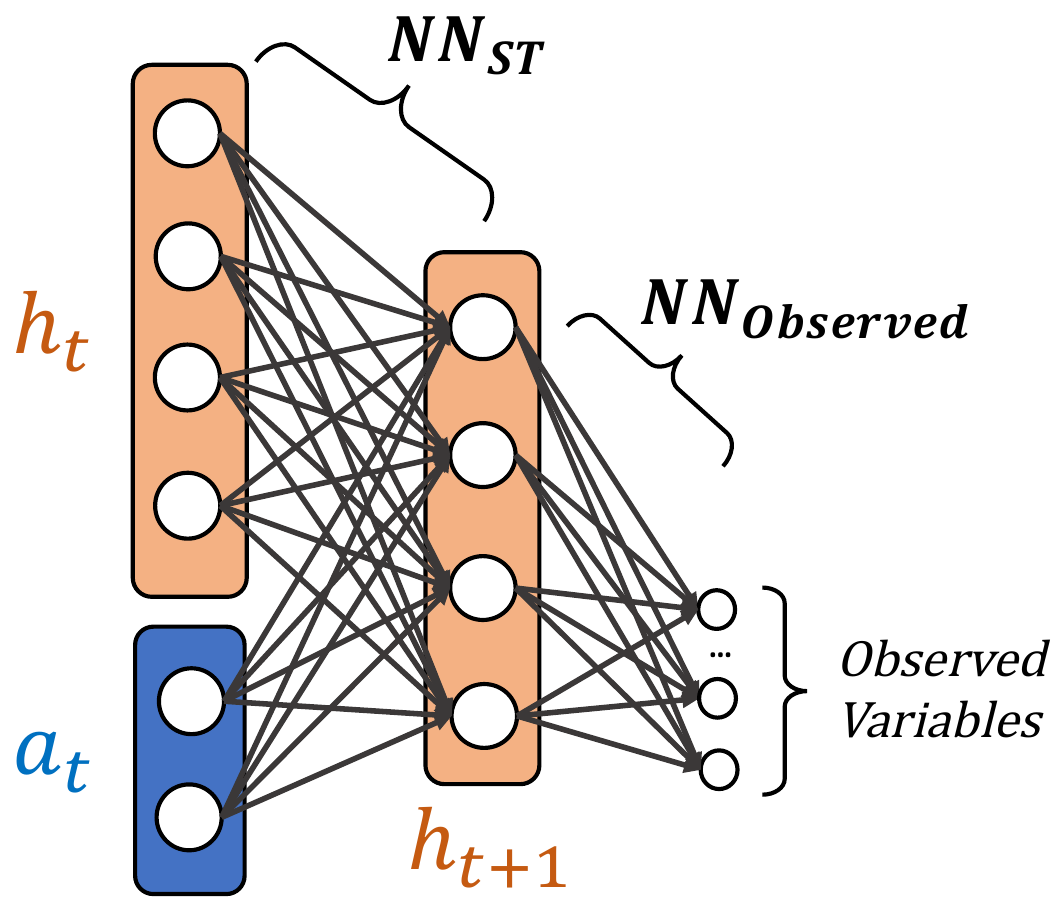}
    \caption{\textbf{Learning the State Transition Function \nameNNST}: Given any $h_t$ and $a_t$, we can extract a representation of a subquery through \nameNNST. We train the function \nameNNST by predicting properties from a set of observed variables. The function \nameNNObserved defines the mapping between the hidden state and these observed variables.}
    \label{fig:nn_statetransition}
\end{figure}

Unfortunately, in the case of query plan enumeration, we cannot easily anticipate the state. The crux of our approach is to represent each state by using a finite dimensional real vector and {\em learn} the state transition function using a deep learning model. To guide the training process for this network, we use input signals and context defined from observed variables that are intimately associated with the state of the database. For example, throughout this work, we use the cardinality of each subquery at any stage of the plan as an observed variable. If a state $h_t$ is represented succinctly with the right amount of information, then we should be able to learn a function, $NN_{observed}$, which maps this state to predicted cardinalities at stage $t$. We show both $NN_{ST}$ and $NN_{observed}$ in ~\autoref{fig:nn_statetransition}.

As we train this model, the parameters of the networks will adjust accordingly based on longer sequences of query operations. With this model, each state will learn to accurately capture a representation. Once trained, we can fix this model and apply reinforcement learning to design an optimal action policy, leading to optimal query plans.



Before describing our approach in more detail, we first introduce fundamental concepts about deep learning and reinforcement learning in the following section.

\section{Background}\label{sec:background}

\textbf{Deep Learning} Deep learning models, also known as feedforward neural networks, are able to approximate a non-linear function, $f$~\cite{Goodfellow:16}. These networks define a mapping from an input $x$ to an output $y$, through a set of learned parameters across several layers, $\theta$. During training, the behavior of the inner layers are not defined by the input data, instead these models must learn how to use the layers to produce the correct output. Since there is no direct interaction between the layers and the input training data, these layers are called \textit{hidden layers}~\cite{Goodfellow:16}. 

These feedforward networks are critical in the context of representation learning. While training to meet some objective function, a neural network's hidden layers can indirectly learn a representation, which could then be used for other tasks~\cite{Goodfellow:16}. In representation learning, there is a trade-off between preserving as much information as possible and learning useful properties about the data. Depending on the output of the network, the context of these representations can vary. It is up to the user to provide the network with enough hints and prior beliefs about the data to help guide the learning~\cite{Goodfellow:16}.

\textbf{Reinforcement Learning}
Reinforcement learning models are able to map scenarios to appropriate actions, with the goal of maximizing a cumulative reward. Unlike supervised learning, the learner (the $agent$) is not explicitly shown which action is best. Instead the agent must discover the best action through trial and error by either exploiting current knowledge or exploring unknown states~\cite{Sutton:16}. At each timestep, $t$, the agent will observe a state of the environment, $s_t$ and will select an action, $a_t$. The action the agent selects depends on the policy, $\pi$. This policy can reenact several types of behaviors. As an example, it can either act greedily or balance between exploration and exploitation through an $\epsilon$-greedy (or better) approach. The policy is driven by the expected rewards of each state, which the model must learn. Given the action selected, the model will arrive at a new state, $s_{t+1}$. At each step, the environment sends the agent a reward, $r_{t+1}$, which signals the ``goodness'' of the action selected. The agent's goal is to maximize this total reward~\cite{Sutton:16}. One approach is to use a value-based iteration technique, where the model records state-action values, $QL(s,a)$. These values specify the long-term desirability of the state by taking into account the rewards for the states that are likely to follow~\cite{Sutton:16}. 




\section{Learning a Query Representation}\label{sec:state_representation}
Given as input a database $D$ and a query $Q$, the first component of our approach is to apply deep learning to derive compact, yet informative representations of queries and the relations they produce. To ensure that these representations are informative, we focus on training these representations to predict subquery cardinalities.

\begin{figure}[t]
    \centering
    \includegraphics[width=0.4\linewidth]{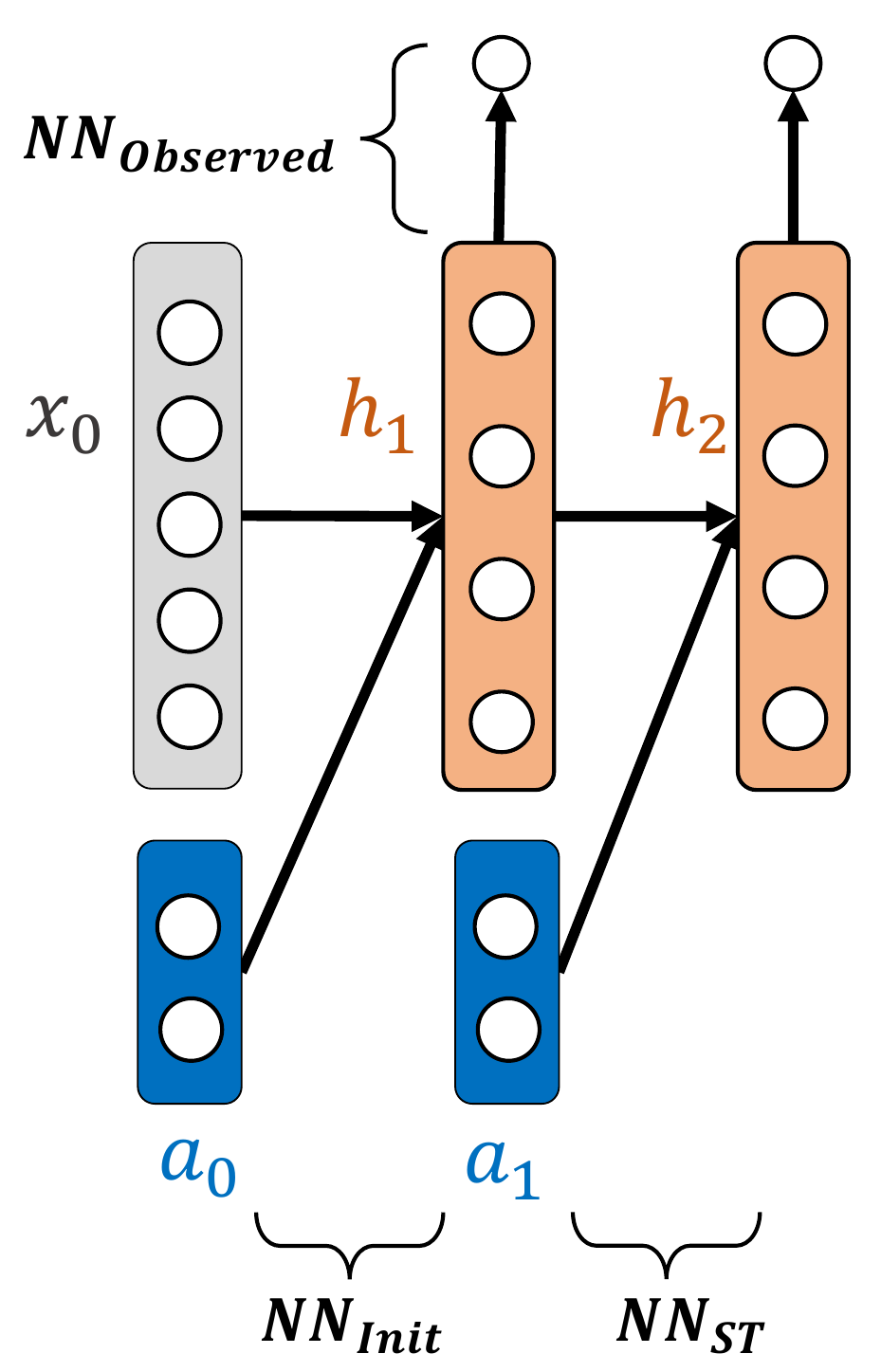}
    \caption{\textbf{ Combined Models \nameNNInit and \nameNNST}}
    \label{fig:combined-models}
\end{figure}

\subsection{Approach}
There are two approaches that we could take. In the first approach, we could transform $(Q,D)$ into a feature vector and train a deep network to take such vectors as input and output a cardinality value. As discussed in the introduction, the problem with this approach is that the size of the feature vector would have to grow with the complexity of databases and queries. This would result in very long, sparse vectors, which would require large training datasets.

Instead, we take a different approach, a recursive approach:
We train a model to predict the cardinality of a query consisting of a single relational operation applied to a subquery as illustrated in \Cref{fig:nn_statetransition}. This model takes as input a pair $(h_t,a_t)$, where $h_t$ is a vector representation of a subquery, while $a_t$ is \textit{a single relational operation on $h_t$}. Importantly, $h_t$ is not a manually specified feature vector, but it is the latent representation that the model learns itself. The \nameNNST function generates these representations by adjusting the weights based on feedback from the \nameNNObserved function. This \nameNNObserved function learns to map a subquery representation to predict a set of observed variables. As we train this model, we use back propagation to adjust the weights for both functions. In this work, we only focus on predicting cardinalities, but we could extend the model to learn representations that enable us to capture additional properties such as more detailed value distributions or features about query execution plans, such as their memory footprint or runtimes.

Before using the recursive \nameNNST model, we must learn an additional function, \nameNNInit, as shown in \Cref{fig:combined-models}. \nameNNInit takes as input ($x_0$,$a_0$), where $x_0$ is a vector that captures the properties of the database $D$ and $a_0$ is a single relational operator. The model outputs the cardinality of the subquery that executes the operation encoded in $a_0$ on $D$. We define the vector, $x_0$ to represent simple properties of the database, $D$. The list of properties we provide next is not definitive and more features can certainly be added. Currently, for each attribute in the dataset $D$, we use the following features to define $x_0$: the $min$ value, the $max$ value, the number of distinct values, and a representation of a one dimensional histogram.

As shown in the figure, we then include the recursive model, \nameNNST, that takes $(h_t,a_t)$ as input and predicts the observed variables of the subqueries as well as the representation, $h_{t+1}$ of the new subquery. We combine these models to train them together. During training, the weights are adjusted based on the combined loss from observed variable predictions. Essentially, we want to learn an $h_1$ representation that captures not only enough information to predict the cardinality of that subquery directly but of other subqueries built by extending it.



\begin{figure}[t]
    \begin{subfigure}[b]{0.25\textwidth}
        \centering
        \includegraphics[width=\linewidth]{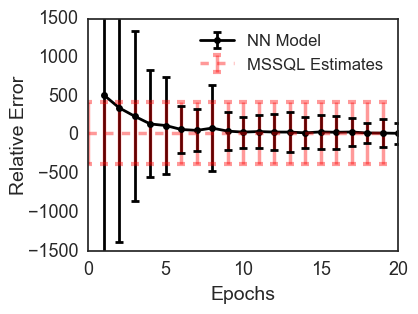}
        \caption{\textbf{Predicting Cardinality ($m=3$)}}
        \label{fig:3col}
    \end{subfigure}~
    \begin{subfigure}[b]{0.25\textwidth}
        \centering
        \includegraphics[width=\linewidth]{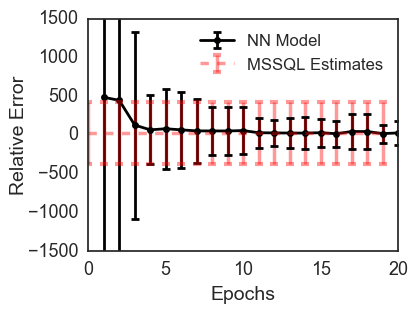}
        \caption{\textbf{Predicting Cardinality ($m=5$)}}
        \label{fig:5col}
    \end{subfigure}
    \caption{Learning $h_1$ for Selection Query}
    \vspace{-0.4cm}
    \label{fig:results1}
\end{figure}

 \begin{figure}[t]
       \begin{subfigure}[b]{0.23\textwidth}
        \centering
        \includegraphics[width=\linewidth]{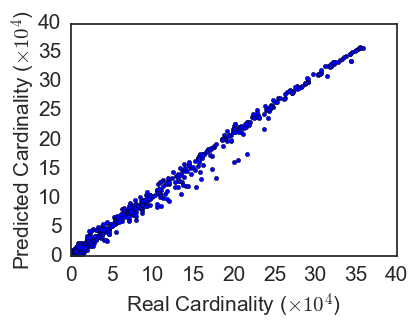}
        \caption{\textbf{Cardinality Predictions for $h_1$}}
        \label{fig:joinh1}
    \end{subfigure}~
    \begin{subfigure}[b]{0.25\textwidth}
        \centering
        \includegraphics[width=\linewidth]{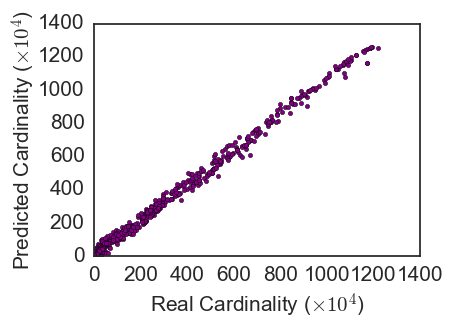}
        \caption{\textbf{Cardinality Predictions for $h_2$}}
        \label{fig:joinh2}
    \end{subfigure}
    \caption{Learning Cardinalities on the Combined Model}
    \vspace{-0.4cm}
    \label{fig:resultsCombined}
\end{figure}

\subsection{Preliminary Results}~\label{sec:results}
We use the publicly available Internet Movie Data Base (IMDB) data set from the Join Order Benchmark (JOB)~\cite{Leis:15}. Unlike TPC-H and TPC-DS, the IMDB data set is real and includes skew and correlations across columns~\cite{Leis:15}. In our experiments, we use Python Tensorflow to implement our approach~\cite{tensorflow2015-whitepaper}.  

\textbf{Training \nameNNInit}: As a first experiment, we initialize $x_0$ with properties of the IMDB dataset and train \nameNNInit to learn $h_1$. $a_0$ represents a conjunctive selection operation over $m$ attributes from the $aka\_title$ relation. We generate 20k queries, where 15k are used for training the model and the rest are used for testing. \nameNNInit contains 50 hidden nodes in the hidden layer. We update the model via stochastic gradient descent with a loss based on relative error and a learning rate of .01.

In \autoref{fig:3col}, we show the cardinality estimation results for selection queries where $m=3$. On the x-axis, we show the number of epochs used during training and on the y-axis we show the relative error with the error bars representing the standard deviation.  We compare our approach $NN Model$  to estimates from SQL Server~\cite{mssql}. We use a commercial engine to ensure a strong baseline. With fewer epochs (less training) the \nameNNInit's cardinality predictions result in significant errors, but at the 6th epoch, the model performs similarly to SQL Server and then it starts to outperform the latter.  

In \autoref{fig:5col}, we increase the number of columns in the selection to $m=5$. In general, we have observed that \nameNNInit takes longer to converge once more columns are introduced. This is expected, as \nameNNInit must learn about more joint distributions across more columns. Nevertheless, the model still manages to improve on SQL Server's estimations by the 9th epoch. 

\textbf{Training \nameNNInit and \nameNNST}: In the previous experiment, we only trained the \nameNNInit model for selection queries over base data. For this next experiment, we predict the cardinality of a query containing both a selection and join operation by using the combined model. Here, $a_0$ represents the selection, while the subsequent action $a_1$ represents the join. Through this combined model, we can ensure that $h_1$ (the hidden state for the selection) captures enough information to be able to predict the cardinality after the join. In \autoref{fig:resultsCombined}, we show the cardinality prediction for $h_1$ and $h_2$. In these scatter plots, the x-axis shows the real cardinality, while the y-axis shows the predicted cardinality from the model. Although there is some variance, $h_1$ was able to hold enough information about the underlying data to make reasonable predictions for $h_2$.

\section{Query Plan Enumeration with Reinforcement Learning}\label{sec:reinforcement_learning}
In this section, we present and discuss our design to leverage the subquery representations from the section above, not only to estimate cardinalities, but to build query plans. Given a query, $Q$, we seek to identify a good query plan by combining our query representations from \nameNNST with reinforcement learning. 

We assume a model-free environment, where transition probabilities between states are not known. At $s_0$, the model only knows about $D$ and $Q$. At this initial state, no query operations have yet taken place. The reinforcement learning agent transitions to a new state by selecting an operation from query $Q$. At each state, we encode an additional contextual vector, $u_t$, which expresses the operations that remain to be done for $Q$. We now describe how to initialize the vector $u_0$ at time $0$:

Given database $D$, we have a set of $n$ relations $\mathcal{R} = \{rel_1,...,rel_n\}$, where each $rel_i$ contains a set of $m$ attributes $\{att_{i_{0}}, ..., att_{i_{m}}\}$. The vector $u_t$ represents a fixed set of equi-join predicates and one-dimensional selection predicates, $\mathcal{C} = \{c_1, ... c_p\}$. We set the $i$-th coordinate in $\mathcal{C}$ accordingly if the corresponding predicate exists in the query $Q$. For example, $c_1$ could represent the following equi-join predicate, $rel_1.att_{1_{0}} = rel_2.att_{2_{3}}$. If this predicate exists in $Q$ we encode it in $u_t$ by updating the value of $c_1$ to 1, otherwise we set it to 0. For selections, we track one-dimensional ranged selections of the following form: $rel_i.att_{i_{j}} <= v$. For now, we allow each attribute to have at most one ranged filter in $Q$. If the selection predicate exists in $Q$, we place the value $v$ in the corresponding element in $\mathcal{C}$. Once the reinforcement agent selects an action (query operation), $a_t$, we can update $u_{t}$ by setting the corresponding element to 0. 

To select good query plans, we need to provide the model with a reward. That reward must either be given at each state or once the entire query plan has been constructed. We have different options. One option is to use the negative cost estimate (computed by the underlying, traditional query optimizer), as the reward for the plan. The system would then learn to mimic that traditional optimizer with the caveat that we currently build only left-deep plans.
A better option that we are currently experimenting with is to use the negative of our
system's cardinality estimates at each step. The limitation, of course, is that this approach only optimizes logical query plans. We plan to extend the reward function in future work to also
capture physical query plan properties. In particular, one approach is to use the negative of the query execution time as the reward. 



Ultimately, the goal of the agent is to discover an optimal policy, $\pi^*$. The policy determines which action the agent will take given the state. As the agent explores the states, the model can update the state-action values, the function $QL(s,a)$, through Q-learning. Q-learning is an \textit{off-policy} algorithm, where it uses two different policies to converge the state-action values~\cite{Sutton:16, silver:15, szepesvari:09}. One is a \textit{behavior} policy which determines which actions to select next. In practice, this is usually an $\epsilon$-greedy policy~\cite{silver:15, Mnih:15}, but other policies can be used as well. The other is the \textit{target} policy, usually a greedy strategy, which determines how values should be updated. 

Initially, all state-action pairs are random values. At each timestep, the agent selects an action (usually based on an $\epsilon$-greedy policy) and observes the reward, $r_{t+1}$ at state $s_{t+1}$. As the agent explores the search space, these state-action pairs will converge to represent the expected reward of the states in future timesteps.

At each state transition, each $QL(s,a)$ is updated as follows:
\begin{equation}
QL(s_t,a_t) \leftarrow QL(s_t,a_t) + \alpha[r_{t+1} + \gamma max_{a'} QL(s_{t+1}, a') - QL(s_{t},a_{t})]
\end{equation}

Where the $max_{a'} QL(s_{t+1}, a')$ represents the maximum value from $s_{t+1}$ given the target greedy policy. We compute the subsequent state given the state transition function, $NN_{ST}$.


\textbf{Open Problems:} Many open problems remain for the above design. As we indicated above, the first open problem is the choice of reward function and its impact on query plan selection.
Another open problem is that the state-space is large even when we only consider selections and join operators as possible actions. Thus, the Q-learning algorithm as initially described is impractical as the state-action values are estimated separately for each unique subquery~\cite{Mnih:15}. In other words, for each query that we train, it is unlikely that we will run into the same \textit{exact} series of states for a separate query. Thus, a better approach is to consider approximate solutions to find values for $QL(s,a)$. We can learn a function, $\hat{QL}(s,a,w)$ to approximate $QL(s,a)$ given parameter weights $w$. This allows the model to generalize the value of a state-action pairs given previous experience with different (but similar) states. This function could either represent a linear function or even a neural network~\cite{silver:15}.

\section{Related Work}\label{sec:relatedwork}
To correct optimizer errors, previous work has used adaptive query processing techniques. Eddies~\cite{Avnur:00} gets rid of the optimizer altogether and instead of building query plan trees, uses an eddy to determine the sequence of operators based on a policy. Tzoumas \ea~\cite{tzoumas:08} took this a step further and transformed it into a reinforcement learning problem where each state represents a tuple along with metadata about which operators still need to be applied and each action represents which operator to run next. 

Leo~\cite{Stillger:01}, was one of the first approaches to automatically adjust an optimizer's estimates. It introduces a feedback loop, allowing the optimizer to learn from past mistakes. This requires successive runs of similar queries to make adjustments. Liu \ea~\cite{Liu:15} uses neural networks to solve the cardinality estimation problem, but primarily focuses on cardinality predictions for selection queries only. Work by Kraska \ea~\cite{Kraska:17} uses a mixture of neural networks to learn the distribution of an attribute to build fast indexes. Instead, our goal is to learn the correlation across several columns and to build query plans. More recently, work by Marcus \ea~\cite{Marcus:18} uses a deep reinforcement learning technique to determine join order for a fixed database. Each state also represents a subquery, but our approach models each state as a latent vector that is learned through a neural network and is propagated to other subsequent states. Their approach uses a policy gradient to determine the best action, while our technique proposes to use a value-based iteration approach. 




\section{Conclusion}\label{sec:discussion} 
In this work, we described a model that uses deep reinforcement learning for query optimization. By encoding basic information about the data, we use deep neural networks to incrementally learn state representations of subqueries. As future work, we propose to use these state representations in conjunction with a reinforcement learning model to learn optimal plans. 

\textbf{Acknowledgements} This project was supported in part by NSF grants IIS-1247469 and Teradata.






\end{sloppypar}

\bibliographystyle{plain}
\bibliography{elasticity}

\end{document}